%% file: 0.Main.tex
\documentclass[10pt]{llncs}
\usepackage{underscore}           
\usepackage{macros}
\usepackage{subfigure}
\usepackage{amsfonts}
\usepackage{graphicx}
\usepackage{epstopdf}
\usepackage{float}
\usepackage{mathtools}
\usepackage[table,xcdraw]{xcolor}
\usepackage{listings}
\usepackage{latexsym}
\usepackage{amssymb}
\newcommand{\remove}[1]{} 
\usepackage{parallel,enumitem}
\usepackage[subfigure]{tocloft}
\usepackage{amsfonts}
\usepackage{graphicx}
\usepackage{cleveref}
\usepackage{soul}
\usepackage{xcolor}
\begin{document}
	
	\title{Controlling Reversibility in Reversing Petri Nets with Application to Wireless Communications}
	\author{Anna Philippou \and Kyriaki Psara }
	\institute{Department of Computer Science,
		University of Cyprus\\
		\email{\{annap,kpsara01\}@cs.ucy.ac.cy} }

	\author{
		Anna Philippou\inst{1},
		Kyriaki Psara\inst{1},
		and Harun Siljak\inst{2}
	}
	\institute{
		Department of Computer Science, University of Cyprus\\
		\email{
		\{annap,kpsara01\}@cs.ucy.ac.cy}
		\and
		CONNECT Centre, Trinity College Dublin, 
		\email
		{harun.siljak@tcd.ie}
	}
	\maketitle
	
	\begin{abstract}
		Petri nets are a  
		formalism for modelling and reasoning about the behaviour of distributed systems. Recently, a reversible approach to Petri nets, Reversing Petri Nets (RPN), has been proposed,
		allowing transitions to
        be reversed spontaneously in or out of causal order. 
        In this work
		we propose an approach for controlling the reversal of actions 
		of an RPN, by associating transitions with conditions whose satisfaction/violation
		allows the execution of transitions in the forward/reversed direction, respectively. 
		We illustrate the 
		framework with a model of a novel, distributed algorithm for antenna selection in distributed 
		antenna arrays. 
	\end{abstract}
	
	\pagestyle{plain}

	\section{Introduction}\label{sec:Introduction}
	\input{1.intro.tex}

	\section{Reversing Petri Nets}\label{sec:ReversingPetriNets}
	\input{2.ReversingPetriNets.tex}

\input{3.semantics.tex}

	\section{Case Study: Antenna Selection in DM MIMO}\label{sec:Case study}

\input{4.CaseStudy.tex}
	\section{Conclusions}\label{sec:Conclusions}
	\input{5.Conclusions.tex}
	\small
	\bibliographystyle{abbrv}

\end{document}

%% file: 1.intro.tex
Reversibility is a phenomenon that occurs in a variety of systems,
e.g., biochemical systems 
and quantum computations. At the same time, it is
often a desirable system property. 
To begin with, technologies based on reversible
computation are considered to be the only way to potentially improve the energy
efficiency of computers beyond the fundamental 
Landauer limit. 
Further applications %
are encountered in programming languages,
concurrent transactions, and fault-tolerant systems, where 
in case of an error a system should reverse back to a safe state. 

As such, reversible computation has been an active topic of research
in recent years and its interplay with concurrency is being investigated
within a variety of theoretical models of computation. 
The notion of causally-consistent reversibility was first introduced 
in the process calculus
RCCS~\cite{RCCS}, advocating that a transition can be undone only if all its effects, if any,
have been undone beforehand. Since then the study of reversibility continued in the context of process calculi~\cite{TransactionsRCCS,Algebraic,LaneseLMSS13,LaneseMS16,CardelliL11},
event structures~\cite{ConRev}, and Petri nets~\cite{PetriNets,RPNs,RPNtoCPN}. 

A distinguishing feature between the cited approaches
is that of {\emph controlling} reversibility: while various
frameworks make no restriction as to when a transition can be reversed (uncontrolled
reversibility), it can be argued that some means of controlling the conditions
of transition reversal is often useful in practice. For
instance, when dealing with fault recovery, 
reversal should only be triggered when a fault is encountered. Based
on this observation, a number of strategies for controlling reversibility have
been proposed: \cite{TransactionsRCCS} introduces the concept of irreversible actions, and \cite{DBLP:conf/rc/LaneseMS12} introduces compensations to deal with  irreversible actions in the context of programming abstractions for distributed systems.  
Another approach for controlling reversibility is proposed in~\cite{ERK} where
an external entity is employed for capturing the order in
which transitions can be executed in the forward or the backward direction. In another line of work,~\cite{LaneseMSS11} defines a roll-back primitive for reversing computation, and in~\cite{LaneseLMSS13}
roll-back  is extended with the possibility of
specifying the alternatives to be taken on resuming the 
forward execution. 
Finally, in~\cite{statistical} the authors associate
the direction of action reversal with energy parameters
capturing environmental conditions of the modelled systems.

In this work we focus on the framework of reversing Petri nets (RPNs)~\cite{RPNs}, which we 
extend with a mechanism for controlling reversibility. This control is enforced with
the aid of conditions associated with transitions, whose satisfaction/violation acts as 
a guard for executing the transition
in the forward/backward direction, respectively.
The conditions are 
enunciated within a simple logical language expressing 
properties relating to available tokens. The mechanism may capture environmental conditions, e.g., changes in temperature, or 
the presence of faults. 
We present a causal-consistent semantics of the framework.
Note that conditional transitions can also be found in existing Petri net models, e.g., in~\cite{CPN}, a 
Petri-net model that associates transitions and arcs with expressions.   

We conclude with the model of a novel antenna selection (AS) algorithm which inspired our framework. Centralized AS in DM MIMO (distributed, massive, multiple input, multiple output) systems \cite{gao2015massive} is computationally complex, demands a large information exchange, and the communication channel between antennas and users changes rapidly. We introduce an RPN-based, distributed, time-evolving solution with reversibility, asynchronous execution and local condition tracking for reliable performance and fault tolerance.

%% file: 2.ReversingPetriNets.tex
In this section we extend the reversing Petri nets of~\cite{RPNs}
by associating transitions with conditions  that control their execution
and reversal, and allow tokens to carry data values of specific types (clauses (2), (6) and (7) in the following definition). We introduce a causal-consistent semantics for the framework. 
\begin{definition}{\rm
		A \emph{\PN}(RPN) is a tuple $(P,T, \Sigma, A, B, F, C, I)$ where:
		\begin{enumerate}
			\item $P$ is a finite set of \emph{places} and
			 $T$ is a finite set of \emph{transitions}.
			\item $\Sigma$ forms a finite set of data types with $V$ the associated
			set of data values.
			\item $A$ is a finite set of \emph{bases} or \emph{tokens} ranged over by $a, b,\ldots$. 
			$\overline{A} = 
			\{\overline{a}\mid a\in A\}$ contains a ``negative" instance for each token and 
			we write ${\cal{A}}=A \cup \overline{A}$.
			
			\item $B\subseteq A\times A$ is a set of undirected \emph{bonds} ranged over by 
			$\beta,\gamma,\ldots$.
			We use the notation $a \bond b$ for a bond $(a,b)\in B$. $\overline{B} = 
			\{\overline{\beta}\mid \beta\in B\}$
			contains a ``negative" instance for each bond and we write ${\cal{B}}=B \cup 
			\overline{B}$.
			\item $F : (P\times T  \cup T \times P)\rightarrow 2^{{\cal{A}}\cup {\cal{B}}}$ 
			is a set of directed labelled \emph{arcs}.
			\item $C:T\rightarrow$ COND is a function that assigns a condition to each 
			transition $t$ such that $type(C(t))=Bool$.
			\item $I : A \rightarrow V$ is a function that associates a 
			data value from $V$ to each token $a$ such that $type(I(a))=type(a)$.
		\end{enumerate}
}\end{definition}

RPNs are built on the basis of a set of \emph{tokens} or \emph{bases} which  correspond to the basic entities that occur in a system. Tokens have a type from the set $\Sigma$, and we write $type(e)$ to denote the type of a token or expression in the language.
Values of these types are associated to tokens of an \RPN via function $I$.
Tokens may occur as stand-alone elements but as computation proceeds they may also merge together to form \emph{bonds}. 
Transitions represent events and are associated with conditions COND 
defined over the data values associated with the tokens
of the model and functions/predicates over the associated data types.
\emph{Places} have the standard meaning. Directed arcs connect places to transitions and vice
versa and are labelled by a subset of ${\cal{A}}\cup {\cal{B}}$. Intuitively, these labels express the requirements for a transition to fire when placed on arcs incoming the transition, and the effects of the transition when placed on the
outgoing arcs. Graphically, a Petri net is a directed bipartite graph where tokens are indicated by $\bullet$, places by circles, transitions by boxes, and bonds by lines between tokens.

 The association of tokens to places is called a \emph{marking}  such that  $M: P\rightarrow 2^{A\cup B}$ where $a \bond b \in M(x)$, for some $x\in P$, implies
$a,b\in M(x)$. 
In addition, we employ the notion of a \emph{history}, which assigns a memory to each
transition 
$H : T\rightarrow 2^\mathbb{N}$. 
Intuitively, a history of $H(t) = \emptyset$ for some $t \in T$ captures that the transition has not taken place, and a history of $k\in H(t)$, 
captures that the transition was executed as the $k^{th}$ transition occurrence and it has not been reversed.
Note that $|H(t)|>1$ may
arise due to cycles in a model. A pair of a marking and a history, $\state{M}{H}$, describes a \emph{state} of a RPN 
with $\state{M_0}{H_0}$ the initial state, where $H_0(t) = \emptyset $ for all $t\in T$. 

We introduce the following notations. We write 
$\circ t =   \{x\in P\mid  F(x,t)\neq \emptyset\}$ and  
$ t\circ = \{x\in P\mid F(t,x)\neq \emptyset\}$
for the incoming and outgoing places of transition
$t$, respectively. Furthermore, we write
$\guard{t}  =   \bigcup_{x\in P} F(x,t)$ 
and $\effects{t}  =   \bigcup_{x\in P} F(t,x)$.
Finally, 
we define $\connected(a,C)$, where $a$ is a token and $C\subseteq A\cup B$ a set of connections,
to be the tokens connected
to $a$ via a sequence of bonds in $B$, together with the bonds creating these connections.

%% file: 3.semantics.tex
In what follows we assume that: (1) transitions do not
erase tokens ($A\cap \guard{t} = A\cap \effects{t}$), and 
(2) tokens/bonds cannot be cloned into more than one outgoing places of a transition
($F(t,x) \cap F(t,y)=\emptyset$ for all $x,y \in P, x\neq y$).  Furthermore, we assume
for all $a\in A, |\multiset{x| a\in M_0(x)}|=1$, i.e., there exists exactly one base of each type in $M_0$. Note that we extend
the exposition of~\cite{RPNs} by allowing transitions to break bonds and by permitting cyclic structures.

\vspace{-0.1cm}
\subsection{Forward execution}
For a transition to be forward-enabled in  an \RPN the following must hold:
\begin{definition}\label{forward}{\rm
		Consider a \RPN $(P,T, \Sigma, A, B, F, C, I)$, a transition $t$,  and a state $\state{M}{H}$. We say that
		$t$ is \emph{forward-enabled} in $\state{M}{H}$ if:
		\begin{enumerate}
			\item If $a\in F(x,t)$  (resp. $\beta\in F(x,t)$)  for some $x\in\circ t$, then $a\in M(x)$ (resp. $\beta\in M(x)$), and if   
			$\overline{a}\in F(x,t)$
		    (resp. $\overline{\beta} \in F(x,t)$)
			for some $x\in\circ t$, then $a\not\in M(x)$ (resp. $\beta\not\in M(x)$), 
			\item If $\beta\in F(t,x)$ for some $x\in t\circ$ and $\beta\in M(y)$ for some $y\in \circ t$ then $\beta\in F(y,t)$,
			\item $E(C(t))$= True. 
		\end{enumerate}
}\end{definition}

Thus, $t$ is enabled in state $\state{M}{H}$ if (1) all tokens and bonds required for the 
transition are available in $t$'s incoming places  and  none of the tokens/bonds whose absence 
is required exists in $t$'s incoming place, 
(2) if a pre-existing bond  appears in an outgoing arc of a transition, then it is also a 
precondition  of the transition to fire, and  
(3) the transition's condition $C(t)$ evaluates to true. We write $E(c)$ for the value of the condition based on the assignment function $I$.

When a transition $t$ is executed in the forward direction, all tokens and bonds occurring in its outgoing arcs are relocated from the input to the output places along with their connected components. The history of $t$ is extended accordingly:

\begin{definition}{\rm \label{forw}
		Given a \RPN $(P,T, \Sigma, A, B, F, C, I)$, a state $\langle M, H\rangle$, and a transition $t$ enabled in $\state{M}{H}$, we write $\state{M}{H}
		\trans{t} \state{M'}{H'}$
		where:
		\[
		\begin{array}{rcl}
		M'(x) & = & 
		M(x)-\bigcup_{a\in F(x,t)}\connected(a,M(x)) \\
		&&\cup \bigcup_{ a\in F(t,x), y\in\circ{t}}\connected(a,M(y)-\guard{t} \cup F(t,x)) 
		\end{array}
		\]
		and 	$H'(t')  =  
		H(t')\cup \{ \max( \{0\} \cup\bigcup_{t''\in T} H(t'')) 
		+1\},$ if  $t' = t $, and 
		$H(t')$, otherwise.
		
}\end{definition}

\subsection{Causal order reversing}
We now move on to  {\em causal-order reversibility}. The following definition enunciates
that a transition $t$ is $co$-enabled (`$co$' standing for causal-order reversing) if it
has been previously executed and all the tokens on the outgoing arcs of the
transition are available in its outplaces. Furthermore, to handle causality in
the presence of cycles, clause (1) additionally requires that all bonds involved in the connected components of such tokens have been constructed by transitions $t'$ that have preceded $t$. Furthermore, clause (2) of the definition requires that
 the condition of  the transition is not satisfied. 

\begin{definition}\label{co-enabled}{\rm
Consider a RPN $(P,T, \Sigma, A, B, F, C, I)$, a state $\state{M}{H}$, and a transition $t\in T$ with $k=\max( H(t))$. Then $t$ is $co$-enabled in  $\state{M}{H}$ if:
	    (1) for all $a\in F(t,y)$ then $a\in M(y)$, and 
	    if  $\connected(a,M(y))\cap \effects{t'} \neq \emptyset$ for some $t'\in T$ with $k'\in H(t')$, then $k'\leq k$, and, 
	    (2) $E(C(t))$= False. 
}\end{definition}

When a transition $t$ is reversed   all tokens and bonds in the
pre-conditions of $t$, as well as their connected components, 
are transferred to $t$'s incoming places. 

\begin{definition}\label{br-def}{\rm
		Given a \RPN 
		a state $\langle M, H\rangle$, and a transition $t$ $co$-enabled in $\state{M}{H}$ with history $k\in H(t)$, we write $ \state{M}{H}
		\rtrans{t} \state{M'}{H'}$
		where:
		\[
		\begin{array}{rcl}
		M'(x) & = & 
		M(x)- \bigcup_{a\in F(t,x)}\connected(a,M(x)) 
		\\ && \cup\bigcup_{ y \in t\circ, a\in F(x,t)}\connected(a,M(y)-\effects{t} \cup F(x,t)) 
		\end{array}
		\]
		
and $H'(t') =  H(t')-\{k\}$ if   $t' = t$, and $H(t')$, otherwise.
		
}\end{definition}

%% file: 4.CaseStudy.tex
\begin{figure}[t]
	\centering
	\subfigure[antennas and users]{\includegraphics[height=3.3cm]{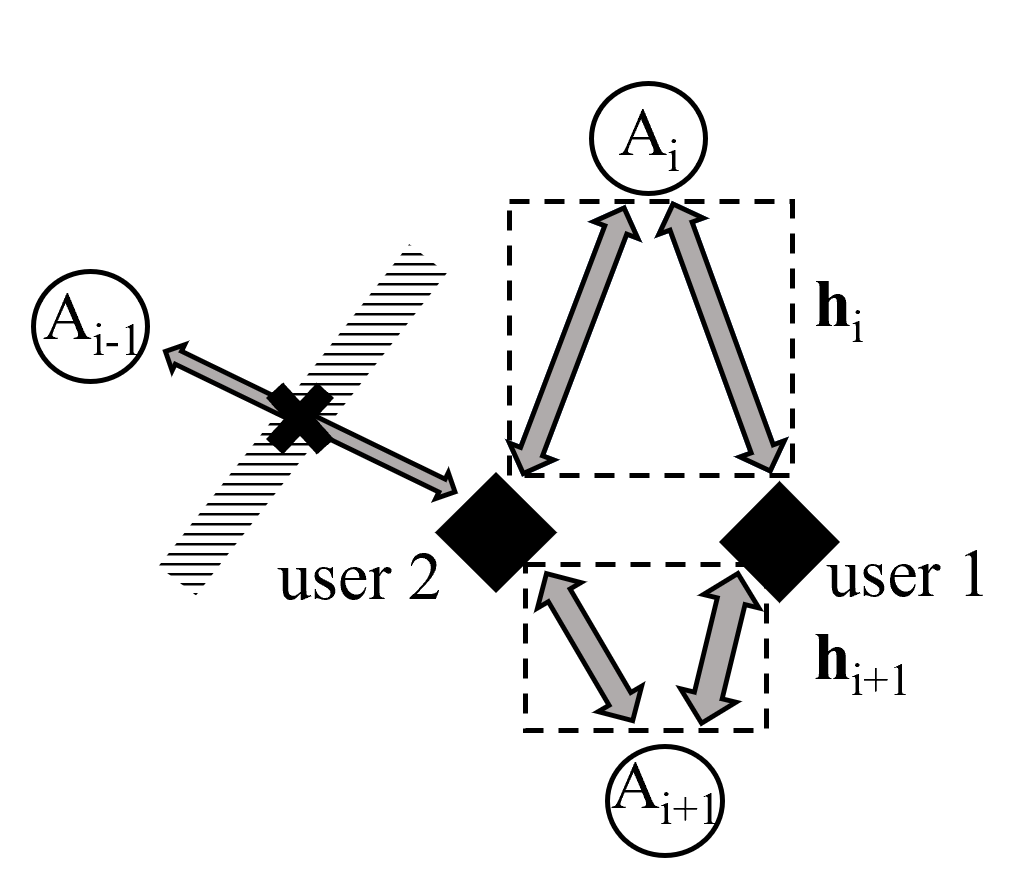}}
	\hspace{2cm}
	\subfigure[a part of the RPN model]{\includegraphics[height=3.3cm]{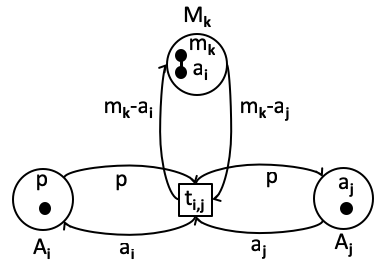}} 
	\caption{RPN for antenna selection in DM MIMO (large antenna array).}
    \label{mechanism}
    \end{figure}

The search for a suitable set of antennas is a sum capacity maximization problem:
 \begin{equation}
\mathcal{C}=\max_{\mathbf{P},\mathbf{H_{c}}}\log_{2}
\det\left(\mathbf{I}+\rho\frac{N_R}{N_{TS}} \mathbf{H_{c}}\mathbf{P}\mathbf{H_{c}}^{H}\right)\label{capac}
\end{equation} 
where $\rho$ is the signal to noise ratio, $N_{TS}$ the number of antennas 
selected from  a total of $N_T$ antennas, $N_{R}$ the number of users, 
$\mathbf{I}$ the $N_{TS}\times N_{TS}$ identity matrix, $\mathbf{P}$ a diagonal
$N_{R}\times N_{R}$ power matrix. $\mathbf{H_{c}}$ is the $N_{TS}\times N_{R}$
submatrix of $N_{T}\times N_{R}$ channel matrix $\mathbf{H}$ \cite{gao2015massive}. Instead of
centralized AS, in our approach (\ref{capac}) is calculated locally for small 
sets of antennas (neighborhoods), switching on only antennas which improve the
capacity: in Fig.~\ref{mechanism}(a), antenna $A_{i-1}$ will not be selected.
In the \RPN interpretation, we present the antennas by places $A_1,\ldots,A_n$, where $n=N_T$, and the  overlapping neighbourhoods
by places $M_1,\ldots,M_h$. These places are connected together via transitions 
$t_{i,j}$, connecting $A_i$, $A_j$ and $M_k$,
whenever there is a connection link between antennas $A_i$ and $A_j$. The transition captures that, based on the neighbourhood knowledge in place $M_k$, antenna $A_i$ may be preferred
over $A_j$ or vice versa (the transition may be reversed). 

To implement the intended mechanism, we employ three types of tokens. 
First, we have the power tokens $p_1,\ldots,p_l$, 
where $l$ is the number of enabled antennas. 
If token $p$ is located on place $A_i$, antenna $A_i$ is considered to be on.
Transfer of these tokens results into new antenna selections, ideally converging
to a locally optimal solution.
Second, tokens $m_1,\ldots,m_h$, each represent one neighborhood. 
Finally, $a_1,\ldots,a_n$, represent the
antennas. The tokens are used as follows:
Given transition $t_{i,j}$ between antenna places $A_i$ and $A_j$ in
neighbourhood $M_k$, transition $t_{i,j}$ is enabled if token $p$ is
available on $A_i$, token $a_j$ on  $A_j$, and bond $(a_i,m_k)$
on $M_k$, i.e.,
$F(A_i,t_{i,j}) = \{p\}$, $F(A_j,t_{i,j})= \{a_j\}$, and
$F(M_k,t_{i,j})=\{(a_i,m_k)\}$. This configuration
captures that antennas $A_i$ and $A_j$ are on and off, respectively.
(Note that the bonds between token $m_k$ and tokens of type $a$
in $M_k$ capture the active antennas in the neighbourhood.)
Then, the effect of the transition
is to break the bond $(a_i,m_k)$, and release token $a_i$ to place
$A_i$, transferring the power token to $A_j$, and creating the bond
$(a_j,m_k)$ on $M_k$, i.e.,
$F(t_{i,j}, A_i) = \{a_i\}$, $F(t_{i,j},A_j)= \{p\}$, and $F(t_{i,j},M_k)
= \{(a_j,m_k)\}$.
The mechanism achieving this for two antennas can be seen in Fig.~\ref{mechanism}(b).

Finally, to capture the transition's condition, an antenna token $a_i$ is associated with data vector $I(a_i) =
\mathbf{h}_i$, $type(\mathbf{h}_i)= \mathbb{R}^2$ ($=\mathbb{C}$), i.e., the 
corresponding row of $\mathbf{H}$. 
The condition constructs the matrix  $\mathbf{H}_c$ of
(\ref{capac}) by collecting the
data vectors $\mathbf{h}_i$ associated with the antenna tokens $a_i$
in place $M_k$: 
$\mathbf{H}_c=(\mathbf{h}_1,...,\mathbf{h}_n)^T$ where
$\mathbf{h}_i=I(a_i)$ if $a_i\in M_k$, otherwise $\mathbf{h}_i=(0\;\ldots\;0)$. 
The transition $t_{i,j}$ will occur if the
sum capacity calculated for all currently active antennas 
(including $a_i$), $\mathcal{C}_{a_i}$, is less than the sum capacity calculated for the same
neighbourhood with the antenna $A_i$ replaced by $A_j$,
$\mathcal{C}_{a_j}$, i.e., $\mathcal{C}_{a_i}<\mathcal{C}_{a_{j}}$. Note that if
the condition is violated, the transition may be executed in the reverse direction.

\remove{
\begin{figure}[t]
	\centering
	\subfigure{\includegraphics[width=12cm]{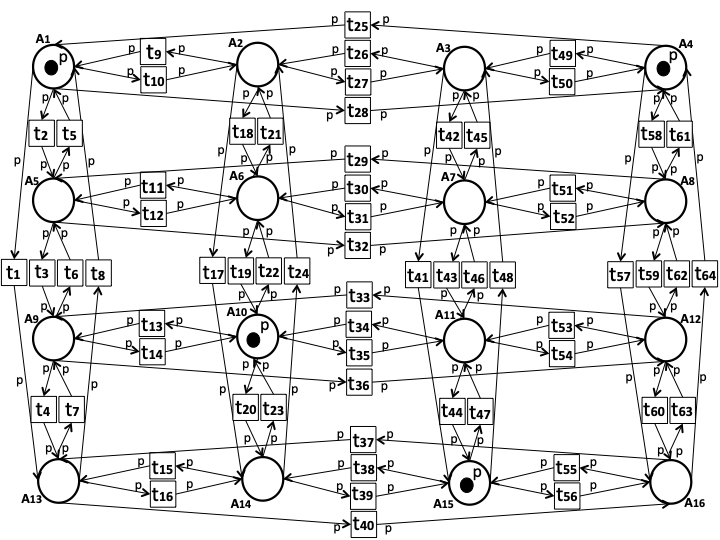}}
	\caption{Antenna selection on massive-MIMO}
	\label{grid}
\end{figure}
}

\remove{
\begin{figure}[t]
	\centering
	\subfigure{\includegraphics[width=7cm]{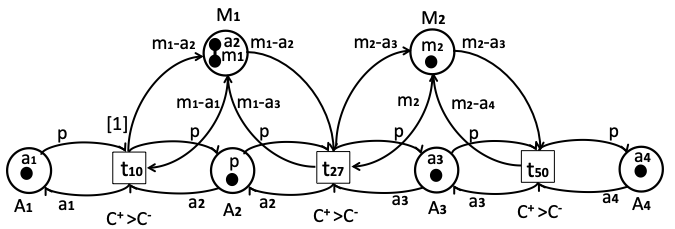}}
	\caption{Memory mechanism on massive-MIMO after the execution of transition $t_{10}$}
    \label{mechanism1}
\end{figure}
}

\begin{figure}[t]
	\centering
	\subfigure{\includegraphics[width=9cm]{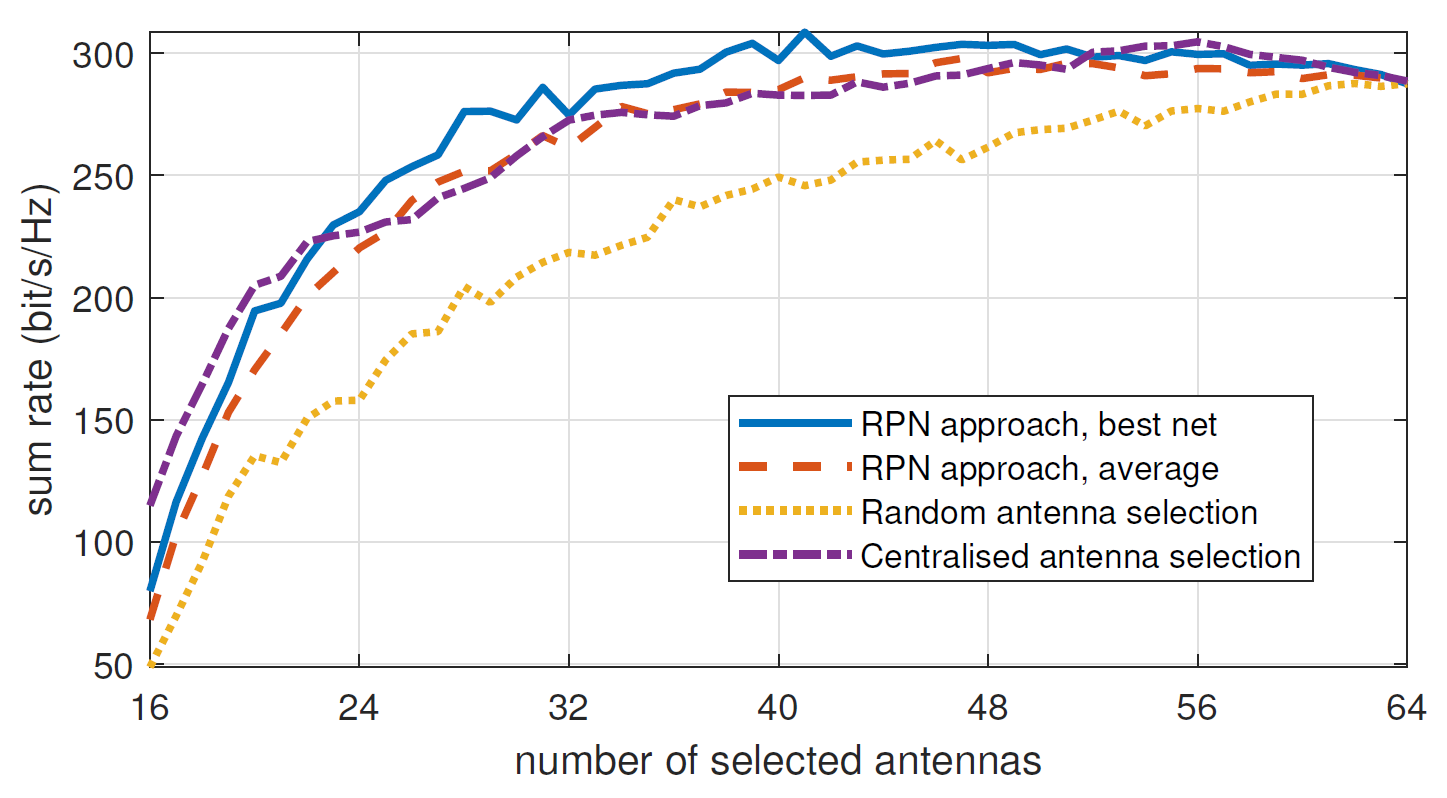}}
	\caption{Results of antenna selection on a distributed 64 antenna array.}
	\label{resant}
\end{figure}

 Results of the RPN-based approach on an array consisting of $64$ antennas serving $16$ users, varying the number of selected antennas from $16$ to $64$ are shown in Fig. \ref{resant} \cite{SPP19}. If we run five RPN models in parallel and select the one with the best performance for the final selection, the results are consistently superior to those of a centralised (greedy) algorithm, and if we run just one (equivalent to the average of the performance of these five models) the results are on par with those of the centralised algorithm.

%% file: 5.Conclusions.tex
We have extended RPNs with conditions 
that control reversibility by determining the direction of transition execution,
and we have applied our framework to model an AS algorithm. Preliminary results show superior performance to centralised approaches. 
Our experience strongly 
suggests that resource management can be  studied and understood 
in terms of RPNs as, along with their visual nature, they offer
a number of relevant features.  
In subsequent work, we plan to extend RPNs for allowing multiple tokens 
of the same base/type to occur in a model
and for developing  out-of-causal-order reversibility semantics in the presence of
conditional transitions as well as the destruction of bonds. 

\noindent{\bf{Acknowledgents:}} This work was partially supported by the European COST Action
IC 1405: Reversible Computation - Extending Horizons of Computing, Science Foundation Ireland (SFI) and
European Regional Development Fund under Grant Number 13/RC/2077, and the EU Horizon 2020
research \& innovation programme under the Marie Sklodowska-Curie grant agreement No 713567.
 